\font\tenrm=cmr10
\font\ninerm=cmr9
\font\nineit=cmti9
\font\ninebf=cmbx9

\font\veinterm=cmr9 scaled 2000
\font\veintebf=cmbx9 scaled 2000

\font\eightrm=cmr8
\font\eightit=cmti8
\font\eightbf=cmbx8

\font\twelvebf=cmbx10 scaled 1200


\def\pmb#1{\setbox0=\hbox{#1}%
   \kern-.025em\copy0\kern-\wd0
   \kern.05em\copy0\kern-\wd0
   \kern-0.025em\raise.0433em\box0}
\def\gta{\mathrel{{\lower 3pt\hbox{$\mathchar"218$}}\hskip-8pt
   \raise 2pt\hbox{$\mathchar"13E$}}}
\def\lta{\mathrel{{\lower 3pt\hbox{$\mathchar"218$}}\hskip-8pt
   \raise 2pt\hbox{$\mathchar"13C$}}}



\def\today{\number\day\space\ifcase\month\or
  January\or February\or March\or April\or May\or June\or
  July\or August\or September\or October\or November\or December\fi
 \space\number\year}
\tolerance=1000

\nopagenumbers
\hsize=6.67truein
\vsize=9truein

\veinterm

\baselineskip=12truept
\noindent{\eightit Journal of Superconductivity, Vol. XXX, 1998}

\vskip .9truein

\noindent{\veintebf 
Tunneling and Photoemission in an SO(6) Superconductor}

\vskip .2in
\twelvebf

\noindent{\bf R.S. Markiewicz$^{a,b}$, C. Kusko$^{a,b,c}$, and M.T. 
Vaughn$^{a}$} 
\vskip .2in
\eightrm

\quad\quad\quad\quad\quad\qquad\quad\quad{\nineit Received 1 June 1998}

\quad\quad\quad\quad\quad\qquad\quad\quad
{\veintebf------------------------------------------------------}

\quad\quad\quad\quad\quad\qquad\quad\quad {\ninerm
Combining the results of tunneling, photoemission and thermodynamic studies, 
the pseudogap}

\quad\quad\quad\quad\quad\qquad\quad\quad {\ninerm
 is unambiguously demonstrated to be caused by Van Hove nesting:
a splitting of the density of} 

\quad\quad\quad\quad\quad\qquad\quad\quad {\ninerm
states peak at $(\pi ,0)$.  The fact that the splitting remains
symmetric about the Fermi level over}

\quad\quad\quad\quad\quad\qquad\quad\quad {\ninerm
 an extended doping range indicates that the 
Van Hove singularity is pinned to the Fermi level.}

\quad\quad\quad\quad\quad\qquad\quad\quad {\ninerm
Despite these positive results, an ambiguity remains as to what instability 
causes the pseudo-}

\quad\quad\quad\quad\quad\qquad\quad\quad {\ninerm
gap.  Charge or spin density waves, superconducting 
fluctuations, and flux phases all remain} 

\quad\quad\quad\quad\quad\qquad\quad\quad {\ninerm
 viable possibilities.  This ambiguity arises because the instabilities of the 
two-dimensional Van}

\quad\quad\quad\quad\quad\qquad\quad\quad {\ninerm
 Hove singularity are
associated with an approximate SO(6) symmetry group, which
 contains }

\quad\quad\quad\quad\quad\qquad\quad\quad {\ninerm
Zhang's SO(5) as a subgroup.  It has two 6-component superspins, one
 of which mixes Zhang's}

\quad\quad\quad\quad\quad\qquad\quad\quad {\ninerm
(spin-density wave plus d-wave superconductivity) superspin with 
a flux phase operator. }

\quad\quad\quad\quad\quad\qquad\quad\quad {\ninerm
 This is the smallest group which can explain striped
 phases in the
cuprates.}

\quad\quad\quad\quad\quad\qquad\quad\quad {\ninerm
Evidence for a prefered hole density in the charged stripes is discussed.}

\quad\quad\quad\quad\quad\qquad\quad\quad 

\quad\quad\quad\quad\quad\qquad\quad\quad
{\veintebf------------------------------------------------------}

\quad\quad\quad\quad\quad\qquad\quad\quad {\ninebf KEY WORDS:} {\eightrm
High-T$_c$ cuprates, underdoped, pseudogap, striped phases.}

\vskip .2truein
\newdimen\fullhsize
\fullhsize=6.67truein \hsize=3.17truein
\def\fullline{\hbox to\fullhsize}

\let\lr=L \newbox\leftcolumn
\output={\if L\lr
   \global\setbox\leftcolumn=\columnbox \global\let\lr=R
  \else \doubleformat \global\let\lr=L\fi
  \ifnum\outputpenalty>-20000 \else\dosupereject\fi}
 \def\doubleformat{\shipout\vbox{\makeheadline
     \fullline{\box\leftcolumn\hfil\columnbox}
     \makefootline}}
  \def\columnbox{\leftline{\pagebody}}
\tenrm

\par\noindent{\bf 1. 11 YEARS of the VAN HOVE SCENARIO:
REMINISCENCES of R.S.M.}
\medskip
Early work on the Van Hove model of high-T$_c$ superconductivity was criticized
for a number of reasons, chiefly (1) the perfect nesting of the Fermi
surface would cause instabilities which would overwhelm superconductivity; and
(2) the Van Hove singularity (VHS) is associated with a special doping, and
hence would require `fine tuning' of the model parameters to  fall at the Fermi
level, whereas superconductivity exists over an extended doping range.  In
June, 1987, I suggested[1] the picture which has become the heart of the 
generalized Van Hove scenario[2].  First, introduction of 
next-nearest-neighbor hopping -- $t^{\prime}$ ($t_{OO}$) in a one (three)-band 
model -- would push the VHS off of half filling, leading to imperfect nesting,
with residual hole pockets and ghost Fermi surfaces[3].  Secondly,
this very sensitivity to instability 
\medskip
\noindent{\veintebf---------}

\noindent{\eightrm 
(a) Physics Department and (b) Barnett Institute, 
Northeastern University, Boston,
MA 02115, U.S.A.  (c)
On leave of absence from Inst.~of Atomic Physics, Bucharest, 
Romania.} 
\eject

\topskip 5.3truein
\par\noindent
could in turn lower the free energy 
precisely when the Fermi surface coincides with the VHS, thereby stabilizing 
this special doping -- and that the reason superconductivity might persist over
such an extended doping range is that the material is inhomogeneous, with one
phase pinned at the VHS.  Due to charging effects, such phase separation would
be nanoscale[4].
\par
While a microscopic model only appeared in 1989[5], I quickly found that 
the doping dependence of both resistivity and
Hall effect could be understood in the context of a percolation 
model[3,6], and Bill Giessen and I noted that the Uemura 
plot[7] could be interpreted in terms of an {\it optimal doping} for
each cuprate, which we identified with the pure VHS phase[8].  At
the 1988 MRS Meeting, Jim Jorgensen asked me if
the phase separation really had to be nanoscale, and I said it was a problem
of the counter\-ions: if the Sr in La$_{2-x}$Sr$_x$CuO$_4$ could be made mobile,
there was no reason that a macroscopic phase separation couldn't occur -- with
one phase pinned at optimal doping.  I didn't know then that Jim and
co-workers had just discovered such a phase in La$_2$CuO$_{4+\delta}$, where the
doping is due to mobile 
\eject
\topskip 0truein
{\bf \ \ }
\vskip .3truein
\par\noindent
interstitial oxygens[9].
\par
One problem was: just where is the VHS.  A VHS at half filling could produce an
antiferromagnetic instability with large pseudogap[10] but no hole
pockets[11], while a VHS at finite doping could explain hole pockets and
optimal doping, but not antiferromagnetism.  I explored this issue by putting
$t^{\prime}$ into a slave boson model, to separate Mott and nesting 
instabilities[5]. I found that the Mott instability 
-- and hence the accompanying antiferromagnetism -- 
remain locked to half filling, while for the
VHS I found a striking surprise: {\it correlation effects pin the Fermi level
close to the VHS over an extended doping range}, extending across the underdoped
regime from half filling to optimal doping (close to the bare VHS).  
Numerous subsequent studies have confirmed the Van Hove pinning (cited on p.
1223 of  Ref. [2]).
\par
In this same paper[5], I demonstrated a viable mechanism for nanoscale phase 
separation: a charge-density wave (CDW) instability is strongly enhanced when
the Fermi level coincides with the bare VHS, leading to a free energy minimum at
optimal doping.  Since the Mott instability produces a separate minimum at half
filling, the energy is lowered by separating into the two end phases.  This
model was clearly distinguished from other models, in that there is a highly
unusual and characteristic fractional hole doping of the hole rich phase (this
doping fixed by the VHS), whereas all other models had a doped phase with one
hole per Cu.  Now I had an {\it embarras de richesses}: both pinning of the VHS
by correlation effects and phase separation, which amounted to a different 
pinning mechanism.  Put another way: were the cuprates characterized by phase
separation or by the (pseudo)gap associated with nesting?  I puzzled over this
issue for a number of years, convinced that somehow both viewpoints were
correct -- rather like the particle vs wave problem in quantum mechanics.  I
was (and am) convinced that both the low-temperature orthorhombic (LTO) and
low-temperature tetragonal (LTT) phases in LSCO was driven by this mechanism,
and I introduced the concept of Van Hove-Jahn-Teller effect to explain the LTO
phase as a dynamic Jahn-Teller (JT) phase[12,13].  But there 
was something missing: there should have been a tetragonal dynamic JT phase at
higher temperatures.  In the years from 1987-1991, I was often asked where was
the evidence for a CDW or other nesting instability.
\par
Things changed when I realized that the `spin gap' and thermodynamic[14] data 
were consistent with a simple model for the pinned Van Hove phase 
\eject
\topskip 0truein
\line{\hfil\bf Markiewicz, et al.}
\vskip .2truein
\par\noindent
[12,15]. 
New photoemission data[16] showed that 
the pseudogap is associated with the band dispersion near $(\pi ,0)$ -- i.e., 
the locus of the VHS, while Tranquada, et al.[17] showed the presence of
nanoscale phase separation in the form of stripes, which appeared to be
suspiciously coextensive with the pseudogap regime.  After writing an extensive
review[2], I put my ideas together into a self-consistent three-band
slave boson calculation, reported at the first Stripes 
Conference[18].  Correlation effects keep the Fermi level close to 
the VHS from half filling to optimal doping.  Near half filling, correlation
effects drive the Cu-O hopping to zero, and the remaining dispersion due to $J$
has a VHS at half filling, and gains additional stability by splitting the VHS
via a flux phase[18,19].  Doping restores the hopping, simultaneously
introducing a strong electron-phonon coupling via modulation of the Cu-O
separation, leading to a maximal CDW instability at optimal doping.  The two
instabilities in turn drive phase separation.  Attempts to model this phase
separation led to good fits to the doping dependence of the photoemission
dispersion.  There was an important prediction: the VHS is found in
photoemission to be below the Fermi level, because it is {\it simultaneously
above the Fermi level}: the pseudogap consists of a splitting of the VHS into
two features at $(\pi ,0)$, but split in energy about the Fermi level.
Photoemission could not reveal the upper VHS, but recent tunneling 
studies[20,21] fully confirm this prediction, as well as demonstrating that the 
lower peak coincides with the photoemission VHS feature.
\par
At the mean-field level, the CDW has long-range order.  However, when 
fluctuations are included in a mode-coupling scheme, there is only short-range
order, with a real pseudogap opening up in the temperature range between the
mean field transition temperature and a much lower transition to long-range
order, driven by interlayer coupling[22].  If this interlayer coupling is
absent, the CDW resembles a {\it quantum critical point} (QCP), with correlation
length diverging as $\xi\sim T^{-1/2}$[22].  It is not a conventional QCP, in
that in the absence of phase sepatation it is not the terminus of a finite
temperature phase transition (i.e., there is no renormalized classical regime).
\bigskip
\par\noindent{\bf 2. SUPERCONDUCTIVITY and TUNNELING}
\medskip
The three-band model[18] revealed that the stri\-ped phases are associated with
two nesting instabilities, flux phase near half filling and CDW at optimal 
doping, leading to a large pseudogap near $(\pi ,0)$ and
\eject
\topskip 0truein
\noindent{\bf SO(6) Superconductor}
\vskip .2truein
\par\noindent
 {\it simultaneously} to
nanoscale phase separation between magnetic and hole-doped stripes.  In the
LSCO system, adding Nd or replacing the Sr with Ba leads to long range stripe
phase order[17] while suppressing superconductivity -- clearly demonstrating
that {\it three distinct phases} must be involved.  In the one-band model[12,15]
this situation is simplified by replacing the striped phase by a uniform CDW
phase.  The key approximation is the {\it Ansatz} of reducing the strong 
correlation effects and phase separation by their most important effect: the
{\it pinning of the Fermi level to the VHS}, over the full doping range from
half filling to optimal doping.  In the overdoped regime, the simplest
approximation is assumed: the band structure stops changing with doping, and
the Fermi level shifts in a rigid band fashion.  
\vskip 2.35truein
{\eightrm Fig.~1. Phase diagram of pinned Balseiro-Falicov model. Circles = net 
tunneling gap, $\Delta_t$.
Inset: Tunneling spectra of a density-wave superconductor, using 
parameters of dotted line in main frame.  Temperatures (from top to bottom) = 
130, 110, 90, 80, 70,
50, and 30K (dashed lines: T above the superconducting transition temperature).}
\vskip0.1truein

The pseudogap seen in tunneling measurements on 
Bi$_2$Sr$_2$CaCu$_2$O$_{8+\delta}$ (Bi-2212)[20,21] can be fit to this model. 
We have calculated the tunneling
spectrum[23], Figure 1, and find a smooth evolution from pseudogap
(CDW) phase to mixed CDW-super\-con\-duct\-or.  For 
a pure CDW, the spectral function is of BCS form:
$$A(k,\omega )=2\pi [u_k^2\delta(\omega -E_{k+})+v_k^2\delta(\omega -E_{k-})],
\eqno(1)$$
with $u_k^2=1-v_k^2=(1+\epsilon_{k-}/ \tilde E_k)/2$,
$E_{k\pm}=(\epsilon_{k+}\pm\tilde E_k)/2$,
$\epsilon_{k\pm}=\epsilon_k\pm\epsilon_{k+Q}$ and $\tilde E_k=\sqrt
{\epsilon_{k-}^2+4G_k^2}$, where the nesting vector $Q=(\pi ,\pi )$, and
the gap $G_k$ and dispersion $\epsilon_k$ are defined in Refs. [12,24].
The model involves three gap parameters, two ($G_0$ and $G_1$) associated with
CDW order, and one $\Delta$ with su-
\eject
\topskip 0truein
{\bf \ \ }
\vskip .25truein
\par\noindent
perconductivity.  Figure 1 shows the 
calculated phase diagram and the net low-T tunneling gap, defined as half the 
peak-peak separation.  The inset shows that in the mixed CDW-superconducting 
state a single gap evolves in the calculated tunneling density of states $\rho 
(E)$ (except for phonon structure).  The ratio of the total gap to the 
CDW/superconducting onset temperature is nearly doping independent, $2\Delta_t
/k_BT^*\simeq 4.1.$  Since we use the model of Balseiro and Falicov (BF)[24] to 
describe the underlying CDW-super\-con\-duct\-iv\-i\-ty competition, we refer to
this as the {\it pinned BF} (pBF) model.  At present, it involves s-wave 
superconductivity, but we are working on an SO(6) generalization, including 
d-wave superconductivity. 
\par
For tunneling along the c-axis into a two-dimen\-sion\-al material, the 
tunneling
density of states (dos) is an average of the in-plane quasiparticle dos[25].
In this case, there is a one-to-one correspondence between features in the 
electronic band dispersion, as measured in photoemission, and peaks in the 
tunneling dos, Fig. 2.  The main tunneling peaks (A) coincide with the split 
electronic energy dispersion near $(\pi ,0)$ and $(0,\pi )$ of the Brillouin 
zone -- and hence with the corresponding photoemission peaks, as found 
experimentally[21].  In the present BCS-like model, the slight discontinuity at 
the phonon energy produces a large peak in the tunneling spectrum (D).  Note
that at $(\pi ,0)$ the CDW and superconducting gaps combine to form a single
feature (A) in both photoemission and tunneling, whereas the gaps split into
two separate features, B associated with superconductivity and C with the CDW,
near $(\pi /2,\pi /2)$.  
\vskip 2.25truein
{\eightrm Fig.~2. Dispersion (a) and tunneling dos (b) near the Fermi level
in the pBF model in the presence of combined CDW and superconducting order.
Parameters are $G_0=$7.45meV, $G_1=$14.3meV, $\Delta =$13.1meV; other parameters
as in Ref.~[23]. Dotted lines = ghost bands.}
\eject
\topskip 0truein
{\bf \ \ }
\vskip .3truein
\par\noindent
The presence of only a single, combined gap at $(\pi ,0)$ is a consequence of an
underlying SO(6) symmetry of the VHS, as discussed in the following section.
In turn, this explains the smooth evolution of the pseudogap into the 
superconducting gap, Fig.~1 insert, which therefore need not be taken as 
evidence for precursor pairing in the pseudogap phase.
\par
A most exciting possibility is that by comparing the photoemission and tunneling
data, one should be able to experimentally {\it measure} the pinning of the 
Fermi level to the VHS.  Indeed, Renner, et al.[20], unaware of this 
prediction[2], have noted that `the pseudogap is centered at the Fermi
level in both under- and overdoped samples.  It is therefore unlikely that the
pseudogap results from a band structure effect.'  
\vskip 2.25truein
{\eightrm Fig.~3. Splitting of the Fermi level from the VHS, measured as 
the normalized difference between the tunneling and photoemission pseudogaps, 
$(\Delta_{PE}-\Delta_{TU})/\Delta_{TU}$.  Solid line = theory in the
absence of pinning; open circles = derived from data of Refs.~[21,26].  Inset:
tunneling dos at several dopings.}
\vskip0.1truein
\par
To quantify the extent of pinning, we test the null hypothesis.  We compare the 
presently available data with the expected tunneling spectra for a pure d-wave
superconductor in the absence of pinning, as doping is varied and the Fermi
level passes through the VHS, Fig.~3.  From the inset, it can be seen that
the two peaks in tunneling are symmetric when the Fermi level is exactly at the 
VHS (optimal doping), while the peak on the photoemission (inverse 
photoemission) side is stronger for underdoped (overdoped) samples, in accord
with experiment.  However, for strong enough doping away from optimal, the peak
should split, which is not seen.  Since the superconducting gap shifts off of
$(\pi ,0)$, we take the difference between the tunneling gap $\Delta_{TU}$ and
the photoemission gap $\Delta_{PE}$ as an experimental measure of the splitting,
Fig.~3.  While there is considerable error, the splitting in overdoped samples
is close to what 
\bigskip
\eject
\topskip 0truein
\line{\hfil\bf Markiewicz, et al.}
\vskip .2truein
\par\noindent
is expected, {\it but the splitting is absent in underdoped
samples} -- strong evidence for Van Hove pinning.  
\bigskip
\par\noindent{\bf 3. SO(6)}
\medskip
A one-dimensional metal is susceptible to a variety of instabilities, including
singlet or triplet superconductivity, CDW's and spin-density waves.  These
instabilities can be organized group-theoretically[27], either on the basis of a
symmetry group, or in terms of a larger spectrum-generating algebra (SGA), which
contains the hamiltonian as a group element, and hence can be used to generate 
the full energy spectrum.  A similar analysis can be applied
to the Van Hove scenario, in terms of an approximate SO(6) symmetry group, or
SO(8) SGA[28].
\par
This SO(6) group contains as subgroups Zhang's SO(5)[29], Yang and Zhang's
SO(4)[30], and Wen and Lee's 
SO(3)[31] (SU(2)).  It includes two 6-dimensional superspins, which form an 
`isospin' doublet[28]: one combines Zhang's SO(5) superspin (antiferromagnetism 
plus d-wave superconductivity) and the flux phase; the other involves s-wave
superconductivity and a CDW (as in the pBF model) with an exotic spin current
phase.  There is a most interesting evolution of these groups from one dimension
to two, Table I.  Lin, et al.[32] analyzed the group structure of a two-leg 
ladder.  They found an SO(8) symmetry group, which involves the SO(6) group as a
subgroup, plus operators which are antisymmetric for $\vec k\rightarrow -\vec 
k$.  These latter operators are essential in the one-dimensional case, where the
Fermi surface consists of two points $\pm k_{Fx}$, but are irrelevant for the 
VHS's, which are on the Brillouin zone boundary, and hence do not couple to 
these operators.  Table I illustrates the evolution of a single superspin (Lin, 
et al.'s d-Mott state) from 1-D to 2-D.  In this
Table, SDW = spin-density wave; sc = superconductivity; $S_{12}^z$ and $P_{12}$
are symbols introduced by Lin, et al.[32] for two of their SO(8) operators,
called band spin difference and relative band chirality; 
and a dash indicates that a corresponding operator is lacking.
Note that the 1-D CDW connects $k_F$ and $-k_F$, while the 2-D CDW discussed
above connects $(\pi ,0)$ and $(0,\pi )$.  Note further that the
SO(6) structure persists down to lad-
\bigskip
\centerline{\bf Table I. Superspin vs. dimensionality}
\vskip 0.1true in
{\settabs 3\columns
\+{\bf 1-D} & {\bf ladder} & {\bf 2-D} \cr
\vskip 0.1true in
\+SDW & SDW & SDW \cr
\+singlet sc & d-wave sc & d-wave sc \cr
\+1-D CDW & $P_{12}$ & --- \cr
\+ --- & $S_{12}^z$ & flux phase \cr
}
\vskip 0.1true in
\eject
\topskip 0truein
\noindent{\bf SO(6) Superconductor}
\vskip .2truein
\par\noindent
ders two cells wide, and hence should
remain valid in describing the striped phases.
\par
In the 1-D metals, the SGA aspect is more fundamental than the symmetry aspect.
The various instabilities are usually not degenerate in energy, as required by a
symmetry group.  Instead, they are governed by the allowed interaction terms, 
$g$'s -- hence the name g-ology -- and the object of research is to derive the
allowed phase diagram as a function of the possible g-values.  In this case, the
SGA is useful in cataloging the allowed instabilities[27].  The situation is 
similar for the VHS.  
The one-band model is not itself symmetric under SO(6), but shows considerable 
signs of the underlying SO(8) SGA.  Thus, the form of the phase diagram in 
Fig. 1 is generic of any competition between a nesting operator and a pairing
operator, while the pseudogap at $(\pi ,0)$ has the simple form 
$$\Delta_t=\sqrt{\Sigma_{i=1}^{12}\Delta_i^2},\eqno(2)$$ 
where the $\Delta_i$'s are the gaps associated 
with the twelve components of both superspin vectors[33].  When $t^{\prime}=0$, 
this vector addition of the gaps holds for the full Fermi surface.  Note that 
for a symmetry group, all the $\Delta_i$'s in Eq. 2 would have equal magnitudes.
\par
The corresponding 2-D G-ology phase diagram can be worked out[28,34], in analogy
with the 1-D case.  In the Hubbard limit, the only interaction is the $U$ term,
and the phase diagram has a natural evolution from SDW at half filling to
d-wave superconductivity in the doped materials.  However, this simple picture 
cannot account for the striped phases, which compete with superconductivity
(e.g., at 1/8 doping, where there is long-ranged stripe order, superconductivity
is suppressed).  By adding a phonon-mediated effective electron-electron
coupling, a CDW phase can be stabilized in the doped material, and competition
between CDW and SDW generates a striped phase.
\bigskip
\par\noindent{\bf 4. POSTSCRIPT: HOW WIDE ARE THE CHARGED STRIPES?}
\medskip
There was considerable discussion at the Conference as to whether the charged
stripes were one cell wide or two cells wide (a.k.a.: site-centered or
bond-centered).  This may seem like a trivial issue, but it can have profound
consequences: the nature of the striped phase at optimal doping, and ultimately
the relevance of the t-J model to high-T$_c$ superconductivity.
\eject
\topskip 0truein
{\bf \ \ }
\vskip .25truein
\par
The issues may be conveniently addressed by considering the White-Scalapino[35]
model of the 1/8 doped state.  The overall magnetic and charge orders are 
consistent with experiment[17], but our main concern here is the actual charge
distribution.  The charge periodicity is four CuO$_2$ cells across, with two
magnetic cells having an average hole density of $\sim 0.07$ holes per Cu, and 
two hole-doped cells, each containing $\sim 0.18$ hole.  Since the magnetic and 
hole-doped stripes have equal width, {\it this is the highest doping at which a
magnetic domain wall model makes sense}.  This is particularly true, since
there is a strong tendency for the magnetic stripes to contain an even number of
cells -- since in that case, there can be a spin gap, as in an even-legged
ladder[36].  Hence, as the doping is increased above 1/8, the magnetic cells can
no longer shrink in width, and there must be a phase transition in the nature
of the stripes.  [If the hole-doped stripes were only one cell wide, this
anomalous behavior would only arise near a doping of 1/4.]
\vskip2.25true in 
{\eightrm Fig.~4. Charged stripes in the tJ model[37], interpreted as 
constant hole density domains of variable width.  Open circles = average hole 
density per row; dashed lines = guides to the eye; solid lines = phase 
separation model.}
\vskip0.1truein
\par
White and Scalapino[37] have indeed reported evidence for such a phase 
transition; here we would like to present a reinterpretation of their data, Fig.
4a of Ref.~[37].  Since the stripes are sensitive to boundary conditions, we
prefer to look at the average hole density along each row (parallel to the
stripes) of the simulation, Fig. 4.  The data (open circles) fall very close to
the form expected for a phase separation model (solid lines), with the same
average densities as at 1/8 filling, but now the magnetic stripes retain their
minimum width, while the hole-doped stripes get wider.  Since 0.18 is close to 
optimal doping, the optimally doped materials are likely to be character-
\eject
\topskip 0truein
{\bf \ \ }
\vskip .3truein
\par\noindent
ized by
a set of widely separated magnetic ladders, with little residual interaction.
The physics will be dominated by the physics of the hole-doped stripes at their
special doping.
\par
This is bad news for the tJ model.  It was specifically designed as a highly
simplified model, which retained just enough physics to accurately describe the
cuprates near the insulating phase at half filling.  It is highly unlikely that
the neglect of the oxygens and electron-phonon interactions will continue to be
valid in the new hole-doped phase. 

We would like to thank A.M. Gabovich for useful discussions about tunneling, and
NATO for enabling him to visit us.  Publication 743 of the Barnett Institute.
\bigskip
\noindent{\bf{REFERENCES}}
\medskip

{\eightrm
\parindent 15pt

\item{1.} R.~S.~Markiewicz, {\eightit Mod.~Phys.~Lett.~B}{\eightbf 1}, 187 
(1987).
\item{2.} R.~S.~Markiewicz, {\eightit J.~Phys.~Chem.~Sol.}, 
{\eightbf 58}, 1179 (1997).
\item{3.} R.~S.~Markiewicz, in ``High~Temperature~Superconductors'', ed.~by
M.~B.~Brodsky, {\eightit et.~al.}, (Pittsburgh,~MRS, 1988), p.~411. 
\item{4.}E.~L.~Nagaev, ``Physics~of~Magnetic~Semiconductors" (Moscow, Mir, 
1983). 
\item{5.} R.~S.~Markiewicz, {\eightit Physica~C}{\eightbf 162-164}, 235 (1989);
{\eightit J.~Phys. Cond.~Matt.} {\eightbf 2}, 665 (1990). 
\item{6.} R.~S.~Markiewicz, {\eightit Physica~C}{\eightbf 153-155}, 1181 (1988).
\item{7.}Y.~J.~Uemura, {\eightit et~al.}, {\eightit Phys.~Rev.~Lett.} {\eightbf 
62}, 2317 (1989).
\item{8.}R.~S.~Markiewicz and B.~C.~Giessen, {\eightit Physica~C}{\eightbf 160},
497 (1989).
\item{9.}J.D. Jorgensen, {\eightit et.~al., Phys.~Rev.~B}{\eightbf 38}, 11337 
(1988).
\item{10.}A.~Kampf and J.~Schrieffer, {\eightit Phys.~Rev.~B}{\eightbf 42}, 7967 
(1990).
\item{11.} W.~Putikka, {\eightit et al.}, to~be~published, {\eightit
J.~Phys.~Chem.~Sol.} {\eightbf 59} (cond-mat/9803141).
\item{12.}R~S.~Markiewicz, {\eightit Physica~C}{\eightbf 193}, 323 (1992).
\item{13.}R~S.~Markiewicz, {\eightit Physica~C}{\eightbf 200}, 65 (1992), 
{\eightbf 210}, 235 (1993), and {\eightbf 207}, 281 (1993).
\item{14.}J.~W.~Loram, {\eightit et.~al., Phys.~Rev.~Lett.} {\eightbf 71}, 1740 
(1993).
\item{15.}R.~S.~Markiewicz, {\eightit Phys.~Rev.~Lett.} {\eightbf 73}, 1310 
(1994).
\vfill
\eject
\topskip 0truein
\line{\hfil\bf Markiewicz, et al.}
\vskip .2truein
\item{16.}D.~S.~Marshall, {\eightit et al.},
{\eightit Phys.~Rev.~Lett.} {\eightbf 76}, 4841 (1996);
A.~G.~Loeser, {\eightit et al.},
{\eightit Science} {\eightbf 273}, 325 (1996);
H.~Ding, {\eightit et al.},
{\eightit Nature} {\eightbf 382}, 51 (1996).
\item{17.}J.~M.~Tranquada, {\eightit et al.},
{\eightit Nature} {\eightbf 375}, 561 (1995).
\item{18.}R.~S.~Markiewicz, {\eightit J.~Supercond.} {\eightbf 10}, 333 (1997);
{\eightit Phys. Rev.~B}{\eightbf 56}, 9091 (1997).
\item{19.}I.~Affleck and J.~B.~Marston, {\eightit Phys.~Rev.~B}{\eightbf 37}, 
3774 (1988);
R.~B.~Laughlin, {\eightit J.~Phys.~Chem.~Sol.} {\eightbf 56}, 1627 
(1995); X.-G.~Wen and P.~A.~Lee, {\eightit Phys.~Rev.~Lett.} {\eightbf 76}, 503 
(1996).
\item{20.}Ch.~Renner, {\eightit et~al., Phys.~Rev.~Lett.} {\eightbf 80}, 149 
(1998).
\item{21.}N.~Miyakawa, {\eightit et~al., Phys.~Rev.~Lett.} {\eightbf 80}, 157 
(1998).
\item{22.}R.~S.~Markiewicz, {\eightit Physica~C}{\eightbf 169}, 63 (1990).
\item{23.}R.~S.~Markiewicz and C.~Kusko, unpublished (cond-mat/\-9802079);
R.S.~Markiewicz, C.~Kusko, and V.~Kidambi, unpublished.
\item{24.}C.~A.~Balseiro and L.~M.~Falicov, {\eightit Phys.~Rev.~B}{\eightbf 
20}, 4457 (1979).
\item{25.}J.~Y.~T.~Wei, {\eightit et~al., Phys.~Rev.~B}{\eightbf 57}, 3650 
(1998).
\item{26.}H. Ding, {\eightit et.~al.,} to~be~published, {\eightit 
J.~Phys.~Chem.~Sol.} {\eightbf 59} (cond-mat/9712100).
\item{27.}A.~I.~Solomon and J.~L.~Birman, {\eightit J. Math. Phys.} {\eightbf 
28}, 1526 (1987).
\item{28.}R.~S.~Markiewicz and M.~T.~Vaughn, to~be~published, {\eightit
J.~Phys.~Chem.~Sol.} {\eightbf 59} (cond-mat/9709137), and {\eightit 
Phys. Rev.~B}{\eightbf 57}, 14052 (1998).
\item{29.}S.-C.~Zhang, {\eightit Science} {\eightbf 275}, 1089 (1997).
\item{30.}C.~N.~Yang and S.-C.~Zhang, {\eightit Mod.~Phys.~Lett.~B}{\eightbf 4},
759 (1990).
\item{31.}X.-G.~Wen and P.~A.~Lee, {\eightit Phys.~Rev.~Lett.} {\eightbf 80}, 
2193 (1998).
\item{32.}H.-H.~Lin, {\eightit et al.},
unpublished (cond-mat/9801285).
\item{33.}C.~Kusko, R.S.~Markiewicz, and M.~T.~Vaughn, unpublished.
\item{34.}H.~J.~Schulz, {\eightit Phys.~Rev.~B}{\eightbf 39}, 2940 (1987).
\item{35.}S.~R.~White and D.~J.~Scalapino, {\eightit Phys.~Rev.~Lett.} {\eightbf
80}, 1272 (1998).
\item{36.}H.~Tsunetsugu, {\eightit et~al. Phys.~Rev.~B}{\eightbf 51}, 16456 
(1995).
\item{37.}S.~R.~White and D.~J.~Scalapino, unpublished (cond-mat/9801274).
}
\vfill
\eject
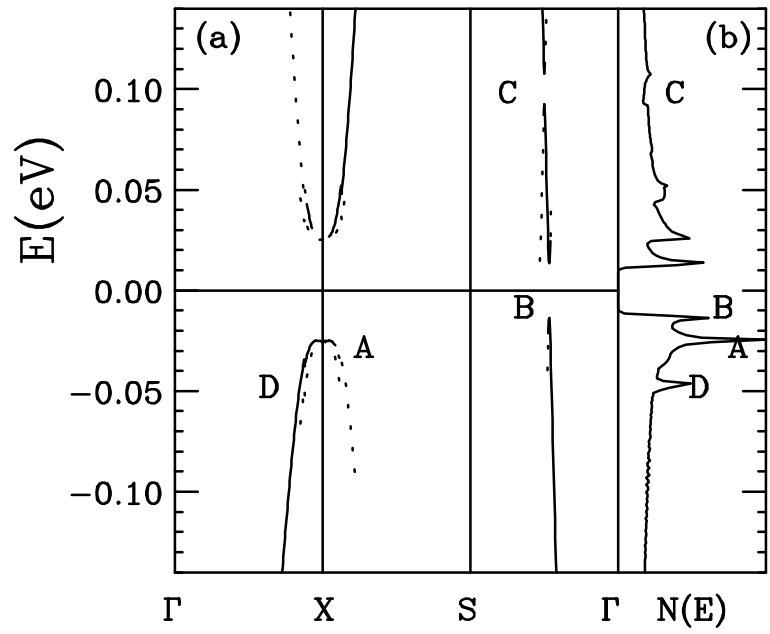
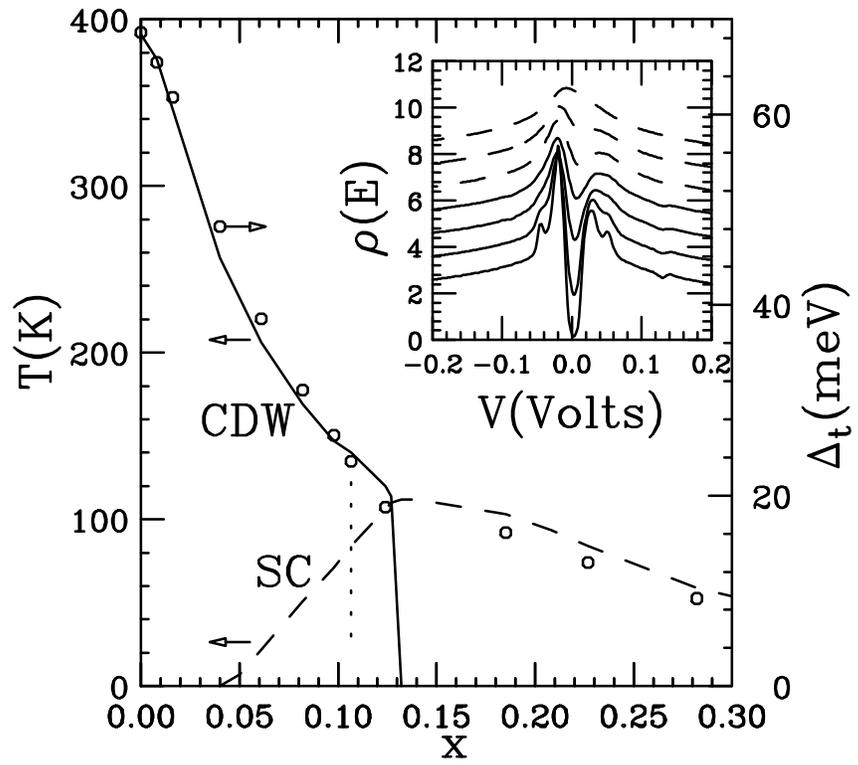
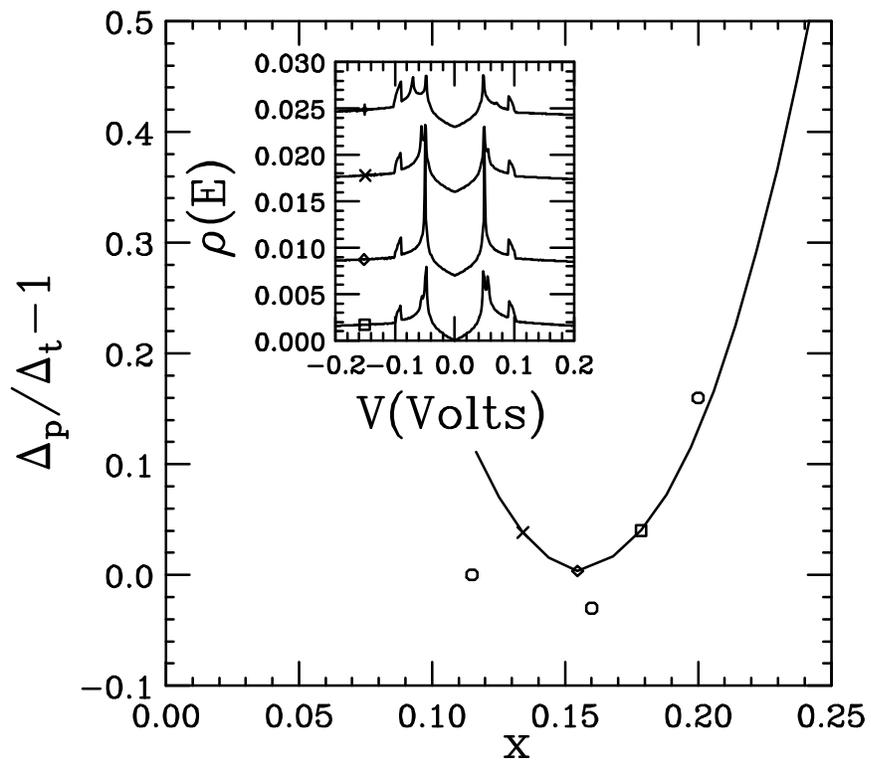
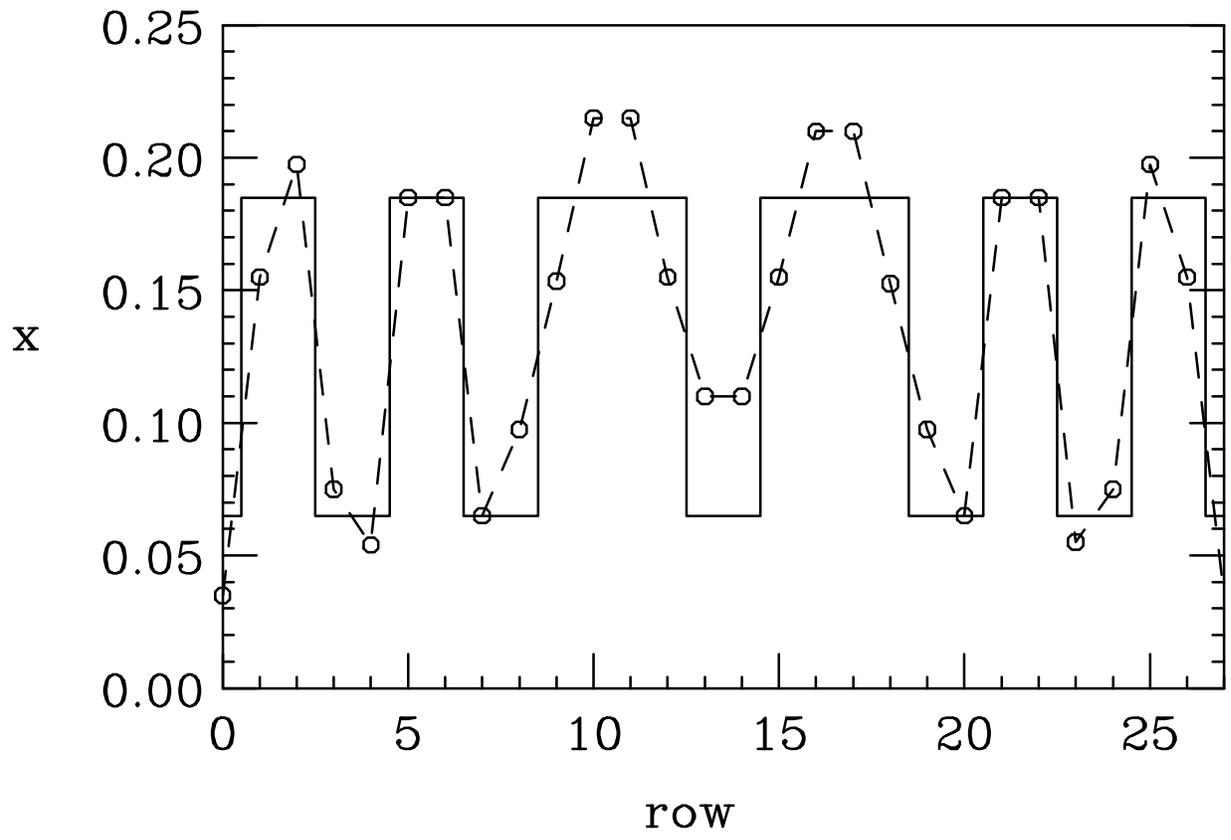
\end